\documentclass[showpacs, showkeys, onecolumn, secnumarabic, aps, pra,
nobibnotes,nofootinbib, floats,superscriptaddress]{revtex4}%
\usepackage{amsmath}
\usepackage{graphicx}
\usepackage{amsfonts}
\usepackage{amssymb}%
\providecommand{\U}[1]{\protect\rule{.1in}{.1in}}
\pdfoutput=1
\providecommand{\U}[1]{\protect\rule{.1in}{.1in}}
\newtheorem{theorem}{Theorem}
\newtheorem{acknowledgement}[theorem]{Acknowledgement}

\begin{document}
\title{Entanglement of two-qubit photon beam by magnetic field}
\author{A. D. Levin}
\email{alexander.d.levin@gmail.com}
\affiliation{Institute of Physics, University of S\~{a}o Paulo, CP 66318, CEP 05314-970,
S\~{a}o Paulo, Brazil}
\author{D. M. Gitman}
\email{gitman@if.usp.br}
\affiliation{P.N. Lebedev Physical Institute, Moscow, Russia; Institute of Physics, USP,
Brazil; Tomsk State University, Russia}
\author{R. A. Castro}
\email{rialcap@hotmail.com}
\affiliation{Institute of Physics, University of S\~{a}o Paulo, CP 66318, CEP 05314-970,
S\~{a}o Paulo, Brazil}

\begin{abstract}
We study the possibility of affecting the entanglement in two-qubit system
consisting of two photons with different fixed frequencies but with two
arbitrary linear polarizations, moving in the same direction, by the help of
an applied external magnetic field. The interaction between the magnetic field
and the photons in our model is achieved through intermediate electrons that
interact both with the photons and the magnetic field. The possibility of
exact theoretical analysis of this scheme is based on known exact solutions
that describe the interaction of an electron subjected to an external magnetic
field (or a medium of electrons not interacting with each other) with a
quantized field of two photons. We adapt these exact solutions to the case
under consideration. Using explicit wave functions for the resulting
electromagnetic field, we calculate the entanglement measures (the information
and the Schmidt ones) of the photon beam as functions of the applied magnetic
field and parameters of the electron medium.

\end{abstract}

\maketitle

\section{Introduction\label{S1}}

Entanglement is a pure quantum property which is associated with a quantum
non-separability of parts of a composite system. Entangled states became a
powerful tool for studying principal questions both in quantum theory and in
quantum computation and information theory
\cite{Bell,Preskill,NieCh00,BellB,OliBoS}. Recently it was proposed a
two-qubit photonic quantum processor that implements two consecutive quantum
gates on the same pair of polarisation-encoded qubits \cite{Barz}. It is
believed that the complete understanding of the nature of quantum entanglement
still requires a detailed consideration of a variety of relatively simple
cases, not only in nonrelativistic quantum mechanics, but in QFT as well. This
explains recent interest to study quantum entanglement and entropy of quantum
states in QFT systems with unstable vacuum, in particular, with strong
external backgrounds that may create particles from the vacuum, see e.g.
\cite{CamPa05,LinChH10,AdaBuP13,BusP13,BruFrF13,FinCa13}. In all these cases
models with exact solutions could be very useful. In the present article, we
are going to use exact solutions of a relativistic quantum mechanical problem
to study how one can affect the entanglement measure of a two-qubit system,
consisting of two photons moving in the same direction with different
frequencies and any of the two possible linear polarizations each, with the
help of an applied external magnetic field. The interaction between the
magnetic field and the photons in the model is performed through intermediate
electrons that interact both with the photons and the magnetic field. An
experimental realization of this theoretical scheme could be the following.
Let us suppose that a beam consisting of the two photons propagating from a
sender to a recipient crosses a region filled with free electrons subjected to
an action of the magnetic field. Thereby, the possibility of creating an
indirect interaction between the external magnetic field and the photon beam
appears. After leaving the region filled with the electrons, the photons will,
in the general case, be registered in an entangled state, if their initial
state was separable, or in an entangled state with a modified initial
entanglement measure, if the initial state was already entangled. The
theoretical support for this scheme is based on exact solutions of quantum
equations of motion that describe the interaction of an electron subjected to
an external magnetic field (or a medium of electrons not interacting with each
other) with a quantized field of two photons with different frequencies and
arbitrary linear polarizations. These exact solutions were studied in Refs.
\cite{164,167,171,178,BagGi90}. In section 2 we apply results of this study to
the case under consideration. In section 3, by using the wave functions
obtained in section 2, we calculate the entanglement measures (the information
and the Schmidt ones) of the photon beam as functions of the applied magnetic
field and parameters of the electron medium. To illustrate the proposed idea,
these calculations are done in the lowest order with respect to a small
parameter, which appears in the problem naturally as a product of the
fine-structure constant and density of the electron medium. It should be noted
that a preliminary consideration of a similar problem was presented in
\cite{Bruno}.

\section{Two-qubit photon beam interacting with an electron placed in constant
uniform magnetic field\label{S2}}

Consider a system of photons (in what follows the photon beam) moving in the
same direction $\mathbf{n}$ and interacting with a Dirac electron. At the same
time whole the system is placed in an external constant and uniform magnetic
field $\mathbf{B}=B\mathbf{n}$ parallel to the photon beam. In fact, this
external field affects directly only the electron, but then, due to the
electron-photon interaction, affects photons as well. The quantum motion of
such a system was studied in detail in Refs. \cite{164,167,171,178,BagGi90}.
Solving the problem in the volume box $V=L^{3}$, one obtains, quantum states
for a photon beam interacting with a free electron gas with a given particle
density $\rho=V^{-1}$.

Below we describe a class of solutions of this kind that correspond to a
photon beam that consists of two photons with different frequencies, each one
with two possible polarizations. The system under consideration is described
by the following Hamiltonian%
\begin{equation}
\hat{H}=\hat{H}_{\gamma}+\gamma^{0}\left(  \boldsymbol{\gamma}\mathbf{\hat{P}%
}\right)  +m\gamma^{0},\;\mathbf{\hat{P}=\hat{p}}+e\left[  \mathbf{\hat{A}%
}(\mathbf{r})+\mathbf{A}^{\mathrm{ext}}(\mathbf{r})\right]  \mathbf{,\ r}%
=\left(  x,y,z\right)  . \label{f1}%
\end{equation}
Here $\hat{H}_{\gamma}$ is the Hamiltonian of the two free transversal
photons, that move in $\mathbf{n}$ direction; $\gamma^{\mu}=\left(  \gamma
^{0},\boldsymbol{\gamma}\ \right)  $ are Dirac gamma matrices \cite{49};
$\mathbf{\hat{A}}(\mathbf{r})$ is the operator-valued vector potential of the
photons in the Coulomb gauge, $\hat{A}_{0}=0$, $\operatorname{div}%
\mathbf{\hat{A}}(\mathbf{r})=0$; $\mathbf{r}$ is the electron coordinate;
$\mathbf{\hat{p}}=-i\boldsymbol{\nabla}$ is the electron momentum operator,
and $\mathbf{A}^{\mathrm{ext}}(\mathbf{r})$ is the vector potential of the
magnetic field in the Landau gauge ($A_{x}^{\mathrm{ext}}=-By,\ A_{0}%
^{\mathrm{ext}}=A_{y}^{\mathrm{ext}}=A_{z}^{\mathrm{ext}}=0$), $B>0$ magnitude
of the magnetic field, $e>0$ is the absolute value of the electron charge, and
$m$ is the electron mass. Following the original work, we represent solutions
in the Heavyside system of units\footnote{where $\hbar=c=1$, and the Coulomb
law takes the form $F=q_{1}q_{2}/4\pi r^{2}$, also $m_{G}=\frac{\hbar}{c}%
m_{H}$,$\ t_{G}=\frac{1}{c}t_{H}$, and$\ e_{G}=\sqrt{\frac{c\hbar}{4\pi}}%
e_{H}$,$\ B_{G}=\sqrt{4\pi c\hbar}B_{H}.$}. Provided that $\mathbf{n}$ is
chosen along the axis $z$, $\mathbf{n}=(0,0,1),$ momenta of the photons from
the beam are%
\begin{equation}
\mathbf{k}_{s}=2\pi L^{-1}\left(  0,0,m_{s}\right)  =\kappa_{s}\mathbf{n}%
,\ s=1,2,\ \kappa_{s}=\kappa_{0}m_{s},\ \kappa_{0}=2\pi L^{-1},\ m_{s}%
\in\mathbb{N}\ , \label{f1a}%
\end{equation}
so that%
\begin{equation}
\hat{H}_{\gamma}=\sum_{s=1,2\mathbf{;}\lambda}\kappa_{s}a_{s,\lambda}%
^{+}a_{s,\lambda},\ \ \mathbf{\hat{A}}(\mathbf{r})=\sum_{s=1,2\mathbf{;}%
\lambda}\frac{1}{e}\sqrt{\frac{\varepsilon}{2\kappa_{s}}}\left[  a_{s,\lambda
}e^{i\kappa_{s}z}+a_{s,\lambda}^{+}e^{-i\kappa_{s}z}\right]  \mathbf{e}%
_{\lambda}\,.\nonumber
\end{equation}
Here $V=L^{3}$ is the quantization box volume, $\mathbf{e}_{\lambda},$
$\lambda=1,2$ are real polarization vectors,\ $(\mathbf{e}_{\lambda}%
\mathbf{e}_{\lambda^{^{\prime}}})=\delta_{\lambda,\lambda^{\prime}}%
$,\ $\left(  \mathbf{ne}_{\lambda}\right)  =0$ and $\varepsilon=e^{2}/L^{3}.$
The photon creation and annihilation operators $a_{s,\lambda}^{+}$ and
$a_{s,\lambda}$ are labeled by $s$ and $\lambda$ and obey the Bose type
commutation relations. The only nonzero relations are%
\begin{equation}
a_{s^{^{\prime}},\lambda^{^{\prime}}}a_{s,\lambda}^{+}-a_{s,\lambda}%
^{+}a_{s^{^{\prime}},\lambda^{^{\prime}}}=\delta_{s,s^{^{\prime}}}%
\delta_{\lambda,\lambda^{^{\prime}}}\,,\ \ s,s^{\prime}=1,2,\ \ \lambda
,\lambda^{\prime}=1,2.\nonumber
\end{equation}
The quantity $\varepsilon$ characterizes the strength of the interaction
between the charge and the plane-wave field. If we interpret $\rho
=V^{-1}=L^{-3}$ as the electron density, then $\varepsilon=e^{2}\rho$. The
dimensionality of $\varepsilon$ is $[\varepsilon]=l^{-3}$ (where $l$ is the
dimensionality of length). Being written with $\hbar$ and $c$ restored, it has
the form:%
\begin{equation}
\varepsilon=\alpha\rho=\frac{\alpha\kappa_{0}^{3}}{8\pi^{3}}\,,\ \ \alpha
=\frac{e^{2}}{\hbar c}=1/137, \label{f4}%
\end{equation}
where $\alpha$ is the fine-structure constant.

Motion of an electron in the magnetic field can be represented as an
oscillator motion, described by new Bose creation, $a_{0}^{+}$, and
annihilation, $a_{0}$, operators,
\begin{equation}
\sqrt{2}a_{0}^{+}=\eta-\partial_{\eta},\;\sqrt{2}a_{0}=\eta+\partial_{\eta
},\ \ \eta=\frac{eBy-p_{x}}{\sqrt{eB}}. \label{f5}%
\end{equation}
The operators $a_{0}$ and $a_{0}^{+}$ commute with every photon operator
$a_{k,\lambda}$ and$\ a_{k,\lambda}^{+}$.

Using a canonical transformation, one can diagonalize the total Hamiltonian
(\ref{f1}) in such a way that it is reduced to two terms that describe two
subsystems -- a subsystem of a quasielectron and a subsystem of quasiphotons
-- that do not interact between themselves,
\begin{align}
&  \hat{H}=\tilde{H}_{\gamma}+\tilde{H}_{e},\ \ \tilde{H}_{e}=r_{0}c_{0}%
^{+}c_{0}+\frac{m^{2}}{2(np)}-\frac{\omega}{2},\nonumber\\
&  \tilde{H}_{\gamma}=\sum_{s=0,1,2;\lambda}r_{s\lambda}c_{s,\lambda}%
^{+}c_{s,\lambda}+\tilde{H}_{\gamma0},\ \ \tilde{H}_{\gamma0}=-\sum
_{s,k=0,1,2;\lambda,\lambda^{\prime}}r_{k\lambda^{\prime}}|v_{s\lambda
,k\lambda^{\prime}}|^{2}+\frac{\epsilon\left(  \kappa_{1}+\kappa_{2}\right)
}{2\kappa_{1}\kappa_{2}}, \label{f6}%
\end{align}
where%
\begin{equation}
\epsilon=\frac{\varepsilon}{(np)}\geq0,\ \ \omega=\frac{eB}{(np)}%
\geq0,\ (np)=p_{0}-p_{z}>0, \label{f9}%
\end{equation}
$p_{0}$ is the electron energy and $p_{z}$ is $z$-projection of the electron
momentum, such that for the electron states $(np)>0$; the quantities
$r_{k\lambda}$ are positive roots of the equation%
\begin{equation}
\sum_{s=1,2}\frac{\epsilon}{r_{k\lambda}^{2}-\kappa_{s}^{2}}=1+\frac{\left(
-1\right)  ^{\lambda-1}\omega}{r_{k\lambda}},\ \ r_{0\lambda}=r_{0}%
\delta_{\lambda1}\ . \label{f12}%
\end{equation}
The matrices $v_{s\lambda,k\lambda^{\prime}}$ are involved in the above
mentioned canonical transformation, which, when written in matrix form, reads%
\begin{align}
&  a=uc-vc^{+},\;a^{+}=c^{+}u^{+}-cv^{+};\ \ c=u^{+}a+v^{T}a^{+},\ c^{+}%
=a^{+}u+av^{\ast};\nonumber\\
&  uu^{+}-vv^{+}=1,\;vu^{T}-uv^{T}=0\,. \label{f10a}%
\end{align}
This linear uniform canonical transformation \cite{225} relates the initial
creation and annihilation operators $a_{k,\lambda}$ and$\ a_{k,\lambda}%
^{+},\ k=0,1,2,$ $a_{0,\lambda}=a_{0}\delta_{\lambda1}$, to the new creation
and annihilation operators $c_{k,\lambda}$ and$\ c_{k,\lambda}^{+}%
,\ k=0,1,2$,\ $c_{0,\lambda}=c_{0}\delta_{\lambda1}$. The free photon
operators $a_{s,\lambda}^{+}$ and $a_{s,\lambda}$,\ $s=1,2$,$\ \lambda=1,2,$
are transformed to new quasiphoton operators $c_{s,\lambda}^{+}$ and
$c_{s,\lambda}$,\ $s=1,2$,$\ \lambda=1,2,$ and the electron creation and
annihilation\ operators $a_{0}^{+}$ and $a_{0}$ are transformed to the
corresponding quasielectron operators $c_{0}^{+}$ and $c_{0}$.

For our purposes it is necessary to write here explicitly only the matrices
$u_{s\lambda,k\lambda^{\prime}}$ and $v_{s\lambda,k\lambda^{\prime}}$ that
correspond to the transformation of the photon operators, i.e., the matrices
with the indices $s,k=1,2$, and$\ \lambda=1,2$. These matrices have the form%
\begin{align}
&  u_{s\lambda,k\lambda^{\prime}}=\left[  \left(  \sqrt{\frac{r_{k\lambda
^{\prime}}}{\kappa_{s}}}+\sqrt{\frac{\kappa_{s}}{r_{k\lambda^{\prime}}}%
}\right)  \frac{\left(  -1\right)  ^{\lambda^{\prime}-1}\delta_{\lambda
,1}-i\delta_{\lambda,2}}{2(r_{k\lambda^{\prime}}^{2}-\kappa_{s}^{2})}\right]
q_{k\lambda^{\prime}},\nonumber\\
&  v_{s\lambda,k\lambda^{\prime}}=\left[  \left(  \sqrt{\frac{r_{k\lambda
^{\prime}}}{\kappa_{s}}}-\sqrt{\frac{\kappa_{s}}{r_{k\lambda^{\prime}}}%
}\right)  \frac{\left(  -1\right)  ^{\lambda^{\prime}-1}\delta_{\lambda
,1}-i\delta_{\lambda,2}}{2(r_{k\lambda^{\prime}}^{2}-\kappa_{s}^{2})}\right]
q_{k\lambda^{\prime}},\nonumber\\
&  q_{k\lambda}=\left[  \frac{\left(  -1\right)  ^{\lambda}\omega}%
{r_{k\lambda}^{3}\epsilon}+2\sum_{s=1,2}(r_{k\lambda}^{2}-\kappa_{s}^{2}%
)^{-2}\right]  ^{-1/2}. \label{f11}%
\end{align}

Stationary states of the system$\ $have the form $\Psi=\Psi_{\gamma}%
\otimes\Psi_{e},$ where $\Psi_{\gamma}$ are state vectors of the
quasiphotons,
\begin{equation}
\Psi_{\gamma}=\prod_{\lambda_{1}=1,2}\frac{(c_{1,\lambda_{1}}^{+}%
)^{N_{1,\lambda_{1}}}}{\sqrt{N_{1,\lambda_{1}}!}}\left\vert 0_{1}\right\rangle
_{c}\otimes\prod_{\lambda_{2}=1,2}\frac{(c_{2,\lambda_{2}}^{+})^{N_{2,\lambda
_{2}}}}{\sqrt{N_{2,\lambda_{2}}!}}\left\vert 0_{2}\right\rangle _{c}\ ,
\label{f14}%
\end{equation}
$c_{s\lambda}\left\vert 0_{s}\right\rangle _{c}=0$,$\ s=1,2$,$\ \forall
\lambda,\ $and $\Psi_{e}$ are state vectors of the quasielectrons, explicit
forms of which are not important for our purposes.

\section{Entanglement in two-qubit photon beam\label{S3}}

Here, to illustrate the proposed idea, we consider the case where the
parameter $\epsilon$ is small. That is why, in what follows, we neglect terms
of the order $o\left(  \epsilon\right)  $ as $\epsilon\rightarrow0.$

In this approximation, for $k=1,2$, we obtain%
\begin{equation}
r_{k\lambda}=\kappa_{k}+\epsilon\frac{\left[  2\kappa_{k}^{2}-\left(
\kappa_{1}^{2}+\kappa_{2}^{2}\right)  \right]  }{(-1)^{\lambda-1}%
2\omega\left[  2\kappa_{k}^{2}-\left(  \kappa_{1}^{2}+\kappa_{2}^{2}\right)
\right]  +\kappa_{k}\left[  5\kappa_{k}^{2}-3\left(  \kappa_{1}^{2}+\kappa
_{2}^{2}\right)  \right]  +\frac{\kappa_{1}^{2}\kappa_{2}^{2}}{\kappa_{k}}}.
\label{f13}%
\end{equation}

\subsection{Photons with antiparallel polarizations\label{SS3.1}}

Consider now the state vector (\ref{f14}) with only two quasiphotons, one of
the first kind, and another of the second kind, and with anti-parallel
polarizations, which we take as $\lambda_{1}=2$ and\ $\lambda_{2}=1$. Such a
state vector corresponds to $N_{1,1}=N_{2,2}=0$ and $N_{1,2}=N_{2,1}=1$ and
has the form
\begin{equation}
\Psi_{\gamma}\left(  \downarrow,\uparrow\right)  =c_{1,2}^{+}c_{2,1}%
^{+}\left\vert 0\right\rangle _{c},\ \ \ \ \left\vert 0\right\rangle
_{c}=\left\vert 0_{1}\right\rangle _{c}\otimes\left\vert 0_{2}\right\rangle
_{c}\ . \label{f15}%
\end{equation}

From the point of view of quasiphotons this is a separable state. However, if
an observer analyzes this state by using tools that register free photons (to
be consistent we have to suppose that in the region where such measurements is
performed the magnetic field and the electron density are zero) he/she will
observe an entangled two free photon state. The corresponding state vector can
be obtained from eq. (\ref{f15}) by expressing all the quasiphoton operators
and vacuum vectors in terms of the corresponding free photon quantities. In
this consideration, we disregard all the terms of the order $o\left(
\epsilon\right)  $. One can easily verify that in this approximation, one can
merely replace the quasiphoton vacuum $\left\vert 0_{s}\right\rangle _{c}$ by
the corresponding free photon vacuum $\left\vert 0\right\rangle _{s}%
$,\ $a_{s,\lambda}\left\vert 0_{s}\right\rangle =0$, so that
\begin{equation}
\Psi_{\gamma}\left(  \downarrow,\uparrow\right)  =c_{1,2}^{+}c_{2,1}%
^{+}\left\vert 0\right\rangle _{c}\simeq c_{1,2}^{+}c_{2,1}^{+}\left\vert
0\right\rangle ,\ \ \left\vert 0\right\rangle =\left\vert 0_{1}\right\rangle
\otimes\left\vert 0_{2}\right\rangle . \label{f16}%
\end{equation}

The operators $c_{1,2}^{+}\ $and $c_{2,1}^{+}$ have to be expressed in terms
of the free photon operators using canonical transformations (\ref{f10a}) and
(\ref{f11}),%
\[
c_{1,2}^{+}=\sum_{s=1,2;\lambda=1,2}a_{s,\lambda}^{+}u_{s\lambda
},\ \ \ c_{2,1}^{+}=\sum_{s=1,2;\lambda=1,2}a_{s,\lambda}^{+}\tilde
{u}_{s\lambda},\ u_{s\lambda}=u_{s\lambda,12},\ \tilde{u}_{s\lambda
}=u_{s\lambda,21}.
\]
Thus,%
\begin{equation}
\Psi_{\gamma}\left(  \downarrow,\uparrow\right)  \simeq\sum_{s,s^{\prime
},\lambda,\lambda^{\prime}}u_{s\lambda}\tilde{u}_{s^{\prime}\lambda^{\prime}%
}a_{s,\lambda}^{+}a_{s^{\prime},\lambda^{\prime}}^{+}\left\vert 0\right\rangle
. \label{f17}%
\end{equation}
Then one can see that in the approximation under consideration we have to
neglect terms of the form $u_{1\lambda}\tilde{u}_{1\lambda^{\prime}%
}a_{1,\lambda}^{+}a_{1,\lambda^{\prime}}^{+}$ and $u_{2\lambda}\tilde
{u}_{2\lambda^{\prime}}a_{2,\lambda}^{+}a_{2,\lambda^{\prime}}^{+}$ in the
right hand side of eq. (\ref{f17}). Thus, we obtain%
\begin{equation}
\Psi_{\gamma}\left(  \downarrow,\uparrow\right)  =\sum_{\lambda,\lambda
^{\prime}}\left[  u_{1\lambda}\tilde{u}_{2\lambda^{\prime}}+u_{2\lambda
^{\prime}}\tilde{u}_{1\lambda}\right]  a_{1,\lambda}^{+}a_{2,\lambda^{\prime}%
}^{+}\left\vert 0\right\rangle . \label{f17a}%
\end{equation}
Introducing the computational basis,
\[
\left\vert \left(  \lambda-1\right)  \left(  \lambda^{\prime}-1\right)
\right\rangle =a_{1,\lambda}^{+}a_{2,\lambda^{\prime}}^{+}\left\vert
0\right\rangle ,\ \left\vert \vartheta_{1}\right\rangle =\left\vert
00\right\rangle ,\ \left\vert \vartheta_{2}\right\rangle =\left\vert
01\right\rangle ,\ \left\vert \vartheta_{3}\right\rangle =\left\vert
10\right\rangle ,\ \left\vert \vartheta_{4}\right\rangle =\left\vert
11\right\rangle ,
\]
we rewrite the vector (\ref{f17a}) as%
\begin{align}
&  \Psi_{\gamma}\left(  \downarrow,\uparrow\right)  =\sum_{\lambda
,\lambda^{\prime}}\left[  u_{1\lambda}\tilde{u}_{2\lambda^{\prime}%
}+u_{2\lambda^{\prime}}\tilde{u}_{1\lambda}\right]  \left\vert \left(
\lambda-1\right)  \left(  \lambda^{\prime}-1\right)  \right\rangle =\sum
_{j=1}^{4}\upsilon_{j}\left\vert \vartheta_{j}\right\rangle ,\ \ \ \sum
_{i=1}^{4}|\upsilon_{i}|^{2}=1,\nonumber\\
&  \upsilon_{1}=u_{11}\tilde{u}_{21}+u_{21}\tilde{u}_{11}=u_{11,12}%
u_{21,21}+u_{21,12}u_{11,21},\nonumber\\
&  \upsilon_{2}=u_{11}\tilde{u}_{22}+u_{22}\tilde{u}_{11}=u_{11,12}%
u_{22,21}+u_{22,12}u_{11,21},\nonumber\\
&  \upsilon_{3}=u_{12}\tilde{u}_{21}+u_{21}\tilde{u}_{12}=u_{12,12}%
u_{21,21}+u_{21,12}u_{12,21},\nonumber\\
&  \upsilon_{4}=u_{12}\tilde{u}_{22}+u_{22}\tilde{u}_{12}=u_{12,12}%
u_{22,21}+u_{22,12}u_{12,21}. \label{f20a}%
\end{align}

The entanglement measure of the system under consideration can be calculated
as the information entropy $E_{I}\left(  \Psi\right)  $ (the von Neumann
entropy with Boltzmann constant the is equal to $1/\ln2$) of the reduced
density operator $\hat{\rho}^{\left(  1\right)  }$ of the subsystem of the
first photon (the same results we obtain calculating the von Neumann entropy
of the reduced operator $\hat{\rho}^{\left(  2\right)  }=\mathrm{tr}_{1}%
\hat{R}$ of the subsystem of the second photon) \cite{Bennet},%
\begin{equation}
E_{I}\left(  \Psi\right)  =-\mathrm{tr}\left(  \hat{\rho}^{\left(  1\right)
}\log_{2}\hat{\rho}^{\left(  1\right)  }\right)  =-\sum_{a=1,2}\lambda_{a}%
\log_{2}\lambda_{a}\ , \label{f21a}%
\end{equation}
where $\lambda_{a},$ $a=1,2,$ are eigenvalues of $\hat{\rho}^{\left(
1\right)  }$. In what follows, we call $E_{I}\left(  \Psi\right)  $ the
information measure. The eigenvalues $\lambda_{a}$ and the information measure
$E_{E}\left(  \Psi\right)  $ have the following form
\begin{align}
&  \lambda_{a}=\frac{1}{2}\left[  \rho_{11}^{\left(  1\right)  }+\rho
_{22}^{\left(  1\right)  }+\left(  -1\right)  ^{a}y\right]  ,\ \ y=\sqrt[+]%
{\left(  \rho_{11}^{\left(  1\right)  }-\rho_{22}^{\left(  1\right)  }\right)
^{2}+4\left\vert \rho_{12}^{\left(  1\right)  }\right\vert ^{2}},\nonumber\\
&  E_{I}\left(  \Psi\right)  =-\frac{1}{\ln4}\left[  \left(  1-y\right)
\ln\left(  \frac{1-y}{2}\right)  +\left(  1+y\right)  \ln\left(  \frac{1+y}%
{2}\right)  \right]  . \label{f22a}%
\end{align}

To obtain the quantity $E_{I}\left(  \Psi\right)  $ in our specific case, we
use the explicit form of the matrices $u$ from eq. (\ref{f11}) and of the
square roots $r_{k\lambda}$ from eq. (\ref{f13}) to calculate the quantities
$\upsilon_{i}$. They latter are%
\begin{equation}
\upsilon_{1}=-(a+b),\ \ \upsilon_{2}=-i(a-b),\ \ \upsilon_{3}=-\upsilon
_{2},\ \ \upsilon_{4}=\upsilon_{1}, \label{f22}%
\end{equation}
where%
\begin{align}
a  &  =\frac{\left(  \sqrt{\frac{\kappa_{1}}{r_{21}}}+\sqrt{\frac{r_{21}%
}{\kappa_{1}}}\right)  \left(  \sqrt{\frac{\kappa_{2}}{r_{12}}}+\sqrt
{\frac{r_{12}}{\kappa_{2}}}\right)  }{4\left(  r_{12}^{2}-\kappa_{2}%
^{2}\right)  \left(  r_{21}^{2}-\kappa_{1}^{2}\right)  \sqrt{\frac{2}{\left(
r_{21}^{2}-\kappa_{1}^{2}\right)  ^{2}}+\frac{2}{\left(  r_{21}^{2}-\kappa
_{2}^{2}\right)  ^{2}}-\frac{\omega}{r_{21}^{3}\epsilon}}\sqrt{\frac
{2}{\left(  r_{12}^{2}-\kappa_{1}^{2}\right)  ^{2}}+\frac{2}{\left(
r_{12}^{2}-\kappa_{2}^{2}\right)  ^{2}}+\frac{\omega}{r_{12}^{3}\epsilon}}%
},\nonumber\\
b  &  =\frac{\left(  \sqrt{\frac{\kappa_{1}}{r_{12}}}+\sqrt{\frac{r_{12}%
}{\kappa_{1}}}\right)  \left(  \sqrt{\frac{\kappa_{2}}{r_{21}}}+\sqrt
{\frac{r_{21}}{\kappa_{2}}}\right)  }{4\left(  r_{12}^{2}-\kappa_{2}%
^{2}\right)  \left(  r_{21}^{2}-\kappa_{1}^{2}\right)  \sqrt{\frac{2}{\left(
r_{21}^{2}-\kappa_{1}^{2}\right)  ^{2}}+\frac{2}{\left(  r_{21}^{2}-\kappa
_{2}^{2}\right)  ^{2}}-\frac{\omega}{r_{21}^{3}\epsilon}}\sqrt{\frac
{2}{\left(  r_{12}^{2}-\kappa_{1}^{2}\right)  ^{2}}+\frac{2}{\left(
r_{12}^{2}-\kappa_{2}^{2}\right)  ^{2}}+\frac{\omega}{r_{12}^{3}\epsilon}}}.
\label{f23}%
\end{align}

Then we obtain the quantity $y$ from eq. (\ref{f22a}) in the approximation
under consideration as%
\begin{align}
&  y\left(  \downarrow,\uparrow\right)  =4\left\vert \left(  a^{2}%
-b^{2}\right)  \right\vert =\left\vert 1-2\frac{\omega}{\epsilon}\left[
\frac{\left(  r_{12}-\kappa_{1}\right)  ^{2}}{\kappa_{1}}-\frac{\left(
r_{21}-\kappa_{2}\right)  ^{2}}{\kappa_{2}}\right]  \right\vert \nonumber\\
&  =1-\epsilon\Phi\left(  \downarrow,\uparrow\right)  ,\ \ 0\leq\epsilon
\Phi\left(  \downarrow,\uparrow\right)  <1,\nonumber\\
&  \Phi\left(  \downarrow,\uparrow\right)  =\omega\frac{\left(  \omega
^{2}\left(  \kappa_{2}-\kappa_{1}\right)  +2\omega\left(  \kappa_{2}%
^{2}+\kappa_{1}^{2}\right)  +\left(  \kappa_{2}^{3}-\kappa_{1}^{3}\right)
\right)  }{2\kappa_{1}\kappa_{2}\left(  \omega-\kappa_{1}\right)  ^{2}\left(
\omega+\kappa_{2}\right)  ^{2}}. \label{f24}%
\end{align}

The asymptotic behavior of the information measure $E_{I}\left(  \Psi_{\gamma
}\left(  \uparrow,\downarrow\right)  \right)  $ as $\epsilon\rightarrow0$
reads
\begin{equation}
E_{I}\left(  \Psi_{\gamma}\left(  \downarrow,\uparrow\right)  \right)
=\frac{\Phi\left(  \downarrow,\uparrow\right)  }{2\ln2}\left[  \epsilon\left(
1-\ln\left(  \Phi\left(  \downarrow,\uparrow\right)  /2\right)  \right)
-\epsilon\ln\epsilon\right]  . \label{f25}%
\end{equation}

We have calculated the entanglement measure for different values of cyclotron
frequencies $\omega=0\div0.5THz$ achievable in a laboratory. For this study we
selected the photon angular frequencies starting with the red light
$\kappa_{1}=2500THz$ and calculated $E_{I}\left(  \Psi_{\gamma}\left(
\downarrow,\uparrow\right)  \right)  $ as a function of $\Delta\kappa
=\kappa_{2}-\kappa_{1}$ ranging from red to ultraviolet. The result is shown
in figure 1 as a surface plot, where the color gradient represents the values
of the entanglement measure.
\begin{figure}[h!]
\centering
\includegraphics[trim=5cm 7cm 5cm 7cm, height=7cm,
width=7cm]{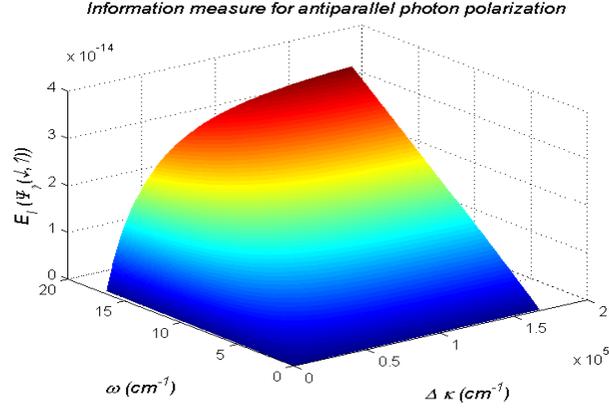}\caption{Information measure $E_{I}\left(  \Psi_{\gamma
}\left(  \downarrow,\uparrow\right)  \right)  $ as a function of $\omega$ and
$\Delta\kappa,~$for $\epsilon=0.1.$}%
\end{figure}

It is also known that for a pure bipartite qudit state one can recognize
entanglement by evaluating the so-called Schmidt measure $E_{S}\left(
\Psi\right)  ,$ which is the trace of the squared reduced density operators
\cite{EisBr01},%
\begin{equation}
E_{S}\left(  \Psi\left(  \downarrow,\uparrow\right)  \right)  =-\mathrm{tr}%
\left[  \left(  \hat{\rho}^{\left(  1\right)  }\right)  ^{2}\right]
=-\sum_{a=1,2}\lambda_{a}^{2}. \label{f25a}%
\end{equation}

The Schmidt measure can be considered as an alternative to the information
entanglement measure. In the case under consideration%
\begin{align}
&  E_{S}\left(  \Psi_{\gamma}\left(  \downarrow,\uparrow\right)  \right)
=1-\left[  \left(  \rho_{11}^{\left(  1\right)  }\right)  ^{2}+\left(
\rho_{22}^{\left(  1\right)  }\right)  ^{2}+2\ \left\vert \rho_{12}^{\left(
1\right)  }\right\vert ^{2}\right] \nonumber\\
&  =1-\left[  2\left(  \left\vert v_{1}\right\vert ^{2}+\left\vert
v_{2}\right\vert ^{2}\right)  ^{2}+8\left\vert \upsilon_{1}\upsilon
_{2}\right\vert ^{2}\right]  =1-16\left(  a^{4}+b^{4}\right)  =2\epsilon
\Phi\left(  \downarrow,\uparrow\right)  . \label{f25c}%
\end{align}

The Schmidt measure calculated for parameters chosen above has quite similar
beheavior as the information measure and is presented on the figure 2
\begin{figure}[h]
\centering
\includegraphics[trim=5cm 7cm 5cm 7cm, height=7cm,
width=7cm]{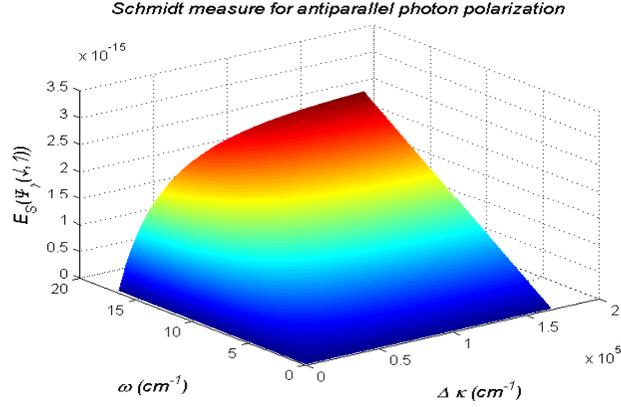}\caption{Schmidt measure $E_{S}\left(
\Psi_{\gamma}\left(  \downarrow,\uparrow\right)  \right)  $ as a function of
$\omega$ and $\Delta\kappa,~$for $\epsilon=0.1.$}%
\end{figure}

\subsection{Photons with parallel polarizations aligned along the magnetic
field\label{SS3.2}}

Let us consider a state (\ref{f14}) with two quasiphotons, one of the first
kind, and another one of the second kind and with parallel polarizations
$\lambda_{1}=1$ and$\ \lambda_{2}=1$. Such a state vector corresponds to
$N_{1,1}=N_{2,1}=1$, $N_{1,2}=N_{2,2}=0$ and has the form
\begin{equation}
\Psi_{\gamma}\left(  \uparrow,\uparrow\right)  =c_{1,1}^{+}c_{2,1}%
^{+}\left\vert 0\right\rangle _{c},\ \ \ \ \left\vert 0\right\rangle
_{c}=\left\vert 0_{1}\right\rangle _{c}\otimes\left\vert 0_{2}\right\rangle
_{c}\ . \label{f26}%
\end{equation}
By the help of the same arguments that were used in the case of anti-parallel
polarizations, we obtain for $\Psi_{\gamma}\left(  \uparrow,\uparrow\right)  $
representation (\ref{f20}) with%
\begin{align}
&  \vartheta_{1}=u_{11,11}u_{21,21}+u_{21,11}u_{11,21},\ \ \vartheta
_{2}=u_{11,11}u_{22,21}+u_{22,11}u_{11,21},\nonumber\\
&  \vartheta_{3}=u_{12,11}u_{21,21}+u_{21,11}u_{12,21},\ \ \vartheta
_{4}=u_{12,11}u_{22,21}+u_{22,11}u_{12,21}. \label{f27}%
\end{align}
Using the explicit form of the matrices $u$ from eq. (\ref{f11}) and square
roots $r_{k\lambda}$ from eq. (\ref{f13}), we can calculate the quantities
$\upsilon_{i}$. They are%
\begin{equation}
\upsilon_{1}=a+b,\ \upsilon_{2}=-i\upsilon_{1},\ \upsilon_{3}=-i\upsilon
_{1},\ \upsilon_{4}=-\upsilon_{1}, \label{f28}%
\end{equation}
where%
\begin{align}
a  &  =\frac{\left(  \sqrt{\frac{r_{11}}{\kappa_{2}}}+\sqrt{\frac{\kappa_{2}%
}{r_{11}}}\right)  \left(  \sqrt{\frac{r_{21}}{\kappa_{1}}}+\sqrt{\frac
{\kappa_{1}}{r_{21}}}\right)  }{4\left(  r_{11}^{2}-\kappa_{2}^{2}\right)
\left(  r_{21}^{2}-\kappa_{1}^{2}\right)  \sqrt{\frac{2}{\left(  r_{11}%
^{2}-\kappa_{1}^{2}\right)  ^{2}}+\frac{2}{\left(  r_{11}^{2}-\kappa_{2}%
^{2}\right)  ^{2}}-\frac{\omega}{r_{11}^{3}\epsilon}}\sqrt{\frac{2}{\left(
r_{21}^{2}-\kappa_{1}^{2}\right)  ^{2}}+\frac{2}{\left(  r_{21}^{2}-\kappa
_{2}^{2}\right)  ^{2}}-\frac{\omega}{r_{21}^{3}\epsilon}}},\nonumber\\
b  &  =\frac{\left(  \sqrt{\frac{r_{11}}{\kappa_{1}}}+\sqrt{\frac{\kappa_{1}%
}{r_{11}}}\right)  \left(  \sqrt{\frac{r_{21}}{\kappa_{2}}}+\sqrt{\frac
{\kappa_{2}}{r_{21}}}\right)  }{4\left(  r_{11}^{2}-\kappa_{1}^{2}\right)
\left(  r_{21}^{2}-\kappa_{2}^{2}\right)  \sqrt{\frac{2}{\left(  r_{11}%
^{2}-\kappa_{1}^{2}\right)  {}^{2}}+\frac{2}{\left(  r_{11}^{2}-\kappa_{2}%
^{2}\right)  {}^{2}}-\frac{\omega}{r_{11}^{3}\epsilon}}\sqrt{\frac{2}{\left(
r_{21}^{2}-\kappa_{1}^{2}\right)  {}^{2}}+\frac{2}{\left(  r_{21}^{2}%
-\kappa_{2}^{2}\right)  {}^{2}}-\frac{\omega}{r_{21}^{3}\epsilon}}}.
\label{f29}%
\end{align}
Then, we obtain the quantity $y$%
\[
y=\sqrt{4\left\vert \upsilon_{1}i\upsilon_{1}+\left(  -i\upsilon_{1}\right)
\left(  -\upsilon_{1}\right)  \right\vert ^{2}}=4\left\vert \upsilon_{1}%
^{2}\right\vert =4\left\vert \left(  a+b\right)  ^{2}\right\vert =1.
\]
One can easily see that in this case, both the information and the Schmidt
entanglement \ measure of the state $\Psi_{\gamma}\left(  \uparrow
,\uparrow\right)  $ are zero%
\[
E_{I}(\Psi_{\gamma}\left(  \uparrow,\uparrow\right)  )=0,\ E_{S}\left(
\Psi_{\gamma}\left(  \uparrow,\uparrow\right)  \right)  =0.
\]

The same result hols true for the state
\[
\Psi_{\gamma}\left(  \downarrow,\downarrow\right)  =c_{1,2}^{+}c_{2,2}%
^{+}\left\vert 0\right\rangle _{c},\ \ \ \ \left\vert 0\right\rangle
_{c}=\left\vert 0_{1}\right\rangle _{c}\otimes\left\vert 0_{2}\right\rangle
_{c}\ .
\]
with two quasiphotons, one of the first kind, and another one of the second
kind and with parallel polarizations $\lambda_{1}=2$,$\ \lambda_{2}=2$,%
\[
E_{I}(\Psi_{\gamma}\left(  \downarrow,\downarrow\right)  )=E_{S}\left(
\Psi_{\gamma}\left(  \downarrow,\downarrow\right)  \right)  =0.
\]

\section{Concluding remarks\label{S4}}

Considering an relevant quantum mechanical model, we have demonstrated that a
two-qubit system that consists of two photons moving in the same direction
with different frequencies and with any of two possible linear polarizations,
can be in a controlled way entangled by applying an external magnetic field in
an electron medium. Then, we succeeded to express the corresponding
entanglement measures (the information and the Schmidt ones) via the
parameters characteristic of the problem, such as photon frequencies,
magnitude of the magnetic field, and parameters of the electron medium. We
have found that, in the general case, the entanglement measures depends on the
magnitude of the applied magnetic field and hence can be controlled by the
latter. As a rule, the entanglement increases with increasing the magnetic
field (with increasing the cyclotron frequency). It should be noted that we
did not consider resonance cases where cyclotron frequency approaches photon
frequencies. Obviously, the entanglement depends on the parameters that
specify the electron medium such as the electron density and electron energy
and momentum. We did not study this dependence in this work; these
characteristics were fixed by choosing a natural, small parameter in our
calculations. The obtained results allow us to see how the entanglement
measures depends on the fixed parameters that characterize the system under
consideration, i.e., on the choice of initial states of the photons and on
photon frequencies. We have come to the following observations: if the both
photon polarizations coincide with themselves and with the direction of the
magnetic field, then no entanglement occurs. The entanglement takes place if
at least one of the photon polarizations is aligned against the direction of
the magnetic field. In this respect, we have a direct analogy with the Pauli
interaction between spin and a magnetic field. However, it seems that this
interaction depends also on the photon frequency and the resulting
entanglement effect depends on the both photon frequencies, or on the first
photon frequency and the difference between the frequencies.

We understand that our study is based on exact solutions of the model problem
- electron interacting with a quantized field of two photons and with a
constant uniform magnetic field. First of all, the constant uniform magnetic
field is an idealization, which cannot be realized experimentally. However,
such an idealization allows exact solutions and is often used in QED.
Sometimes, one can verify that local space-time processes do not depend
essentially on the asymptotic behaviour of the external field. More realistic
results could be obtained if magnetic fields vanishing on space-time infinity
were used. In calculating the entanglement measures, we have also used the
first-order approximation in the natural parameter $\epsilon=\frac{\rho
}{137(np)}$, supposing that it is small, $\epsilon\sim0,1.$ In fact, this
imposes restrictions on the electron density $\rho$ and electron energy and
momentum, $(np)=p_{0}-p_{z}.$ However, this approximation was enough for our
semi qualitative preliminary study of the problem. All the above-mentioned
approximations would have to be carefully estimated in order to that real
numbers for possible experimental realization of the controlled entanglement
of photons by a magnetic field might be extracted.

Finally, we stress that the principle aim of our study was to demonstrate a
possibility for entangling photon beams by the help of an external magnetic
field. In contrast with the well-known possibilities of doing this by using
crystal devices, the present way allows one to change easily and continuously
the entanglement measure. We do not discuss in the article how one can
experimentally measure the entanglement measure. This is an important general
problem, which, however, is beyond the scope of our study. In each specific
case, it demands significant efforts to realize such experiments. Important
results in this direction can be found in Refs.
\cite{AdaBuP13,BusP13,BruFrF13,FinCa13}. Description of some experiments that
use NMR methods and allow one to detect quantum entanglement can be found in
the book \cite{OliBoS} (see section 6.3 there).

\begin{acknowledgement}
Gitman is grateful to the Brazilian foundations FAPESP and CNPq for permanent
support, the work is partially supported by the Tomsk State University
Competitiveness Improvment Program. Castro is grateful to the Brazilian
foundation CNPq for its support.
\end{acknowledgement}

\end{document}